\documentclass[aps, a4paper, superscriptaddress, nofootinbib,  twocolumn, 10pt]{revtex4}
\usepackage{epsfig}
\usepackage{graphicx}
\usepackage{slashed}
\usepackage{mathtools}

\newcommand{\beq}{\begin{equation}}
\newcommand{\eeq}{\end{equation}}
\newcommand{\bea}{\begin{eqnarray}}
\newcommand{\eea}{\end{eqnarray}}
\newcommand{\ba}{\begin{align}}
\newcommand{\ea}{\end{align}}
\newcommand{\bfig}{\begin{figure}}
\newcommand{\efig}{\end{figure}}

\newcommand{\gev}{\, \text{GeV}}

\newcommand{\tin}{t_{\text{in}}}

\begin{document}

\phantom{}
\vspace*{-17mm}

\title{Pion form factor and low-energy hadronic contribution to muon $g-2$ by
analytic extrapolation: consistency and sensitivity tests}

\author{B.Ananthanarayan}
\affiliation{Centre for High Energy Physics,
Indian Institute of Science, Bangalore 560 012, India}
\author{Irinel Caprini}
\affiliation{Horia Hulubei National Institute for Physics and Nuclear Engineering,
P.O.B. MG-6, 077125 Magurele, Romania}
\author{ Diganta Das}
\affiliation{Department of Physics and Astrophysics, University of Delhi, Delhi 110007, India}



\begin{abstract} The largest error in the  theoretical determination of the  muon anomalous magnetic moment is due to the low-energy hadronic 
vacuum polarization, which cannot be calculated by perturbative QCD and requires nonperturbative techniques. 
 Recently, an accurate determination of  the low-energy two-pion contribution to muon $g-2$ has been obtained by a parametrization-free analytic continuation of the pion vector form factor from other kinematical regions. In this work we compare the results of the analytic continuation with direct determinations  at low momenta from experiment and lattice QCD.  We also explore the sensitivity of the method to the timelike data on the modulus of the form factor used as input, by extending the input region to energies up to 0.76 GeV.

\end{abstract}
\maketitle
\section{Introduction}

The  muon magnetic anomaly $a_\mu=(g-2)_\mu/2$ plays a special role in the history of elementary particle physics.
Its value at one-loop order in quantum electrodynamics (QED) calculated by Julian Schwinger \cite{Schwinger}
earned him the Nobel Prize in physics.  Today, this quantity turns out to provide a crucial test of the Standard
Model (SM) of particle physics.
The present experimental value 
\beq
a_\mu^{\text{exp}} = (11 659 209.1\pm 5.5_{\text{stat}}\pm 3.3_{\text{sys}}) \times 10^{−10}
\eeq
is dominated by the Brookhaven measurement \cite{Bennett:2006fi}.  
On the theoretical side, $a_\mu$ has been calculated in the SM to very high orders in QED and by semi-phenomenological approaches for the hadronic contribution (for recent reviews see \cite{Davier:2017,  FJ:Springer, Teubner:2018}). The most recent value obtained in \cite{Teubner:2018} 
\beq
a_\mu^{\text{SM}} = (11 659 182.04 \pm 3.56) \times 10^{−10}
\eeq
confirmed the $\sim 3.7 \sigma$ deviation from the experimental value already known since some time. This discrepancy, if confirmed by future investigations, would  be of crucial significance, being a serious indirect indication of the existence of new physics beyond the SM.

Efforts are currently underway  to improve the experimental measurements by the
``Muon $g-2$  E989 Experiment''  at Fermilab \cite{Venanzoni:2012qa} and the proposed future J-PARC experiment \cite{JPARC}.
In parallel, the efforts  of the ``Muon $g-2$ Theory Initiative'' \cite{Theory} and the groups involved within it are devoted to the improvement of the calculation of $a_\mu^{\text{SM}}$ and of its uncertainty.

 At present, the largest theoretical error of $a_\mu^{\text{SM}}$ is due to the low-energy hadronic radiative corrections described by hadronic vacuum polarization and hadronic
light-by-light-scattering Feynman diagrams. These corrections cannot 
be evaluated by quark and gluon loops in perturbative quantum chromodynamics (QCD) and require the application of nonperturbative techniques. 
 The hadronic vacuum polarization brings the dominant contribution and its accurate knowledge is necessary in order to control the theoretical
uncertanties. The hadronic light-by-light scattering is less significant quantitatively but more difficult to evaluate. In this work we consider the low-energy contribution to the muon $g-2$ of the hadronic vacuum polarization. 

 The main contribution to this hadronic correction is given by the two-pion states and is expressed by unitarity as a weighted integral over of the c.m.s. energies of the cross section $\sigma (e^+e^-\to\pi^+\pi^-)$ of the annihilation of $e^+e^-$ into a pair of charged pions. Several high-statistics $e^+e^-$ experiments \cite{Akhmetshin:2003zn}-\cite{Ablikim:2015orh} have been designed and operated recently in order to increase the precision of $a_\mu$ determination. However, the experiments are difficult at low energies, below $0.6 \gev$,  where recent measurements have been performed only by two experiments, BABAR \cite{Aubert:2009ad,Lees:2012cj} and KLOE \cite{Ambrosino:2008aa,Ambrosino:2010bv,Babusci:2012rp}. 

The direct evaluation based on experimental data can be avoided by expressing  the two-pion  contribution to $a_\mu$ as \cite{Davier:2017}
\begin{equation} \label{eq:amu}
a_\mu^{\pi\pi}= \frac{\alpha^2 m_\mu^2}{12 \pi^2}\int_{4 m_\pi^2}^{\infty} \frac{dt}{ t} K(t)\ \beta^3_\pi(t) \  F_{\rm FSR}(t)\
|F_\pi^V(t)|^2,\end{equation} 
in terms of the  pion electromagnetic form factor $F_\pi^V(t)$, defined by the matrix element of the current operator between charged pion states:
\beq\label{eq:def} 
 \langle \pi^+(p')\vert J_\mu^{\rm elm}(0) \vert
\pi^+(p)\rangle= (p+p')_\mu F_\pi^V(t).
\eeq 
 The remaining factors in (\ref{eq:amu}) are: the phase-space factor $\beta_\pi(t)=(1-4 m_\pi^2/t)^{1/2}$, the QED kernel  \cite{FJ:Springer}
\begin{equation} \label{eq:K}
K(t) = \int_0^1 du(1-u)u^2(t-u+m_\mu^2u^2 )^{-1}
\end{equation}
and  a final-state radiation (FSR) correction
\beq
 F_{\rm FSR}(t)=\left(1+\frac{\alpha}{\pi}\,\eta_\pi(t)\right),
\eeq
  calculated usually in scalar QED \cite{Davier:2017, FJ:Springer}. Finally, $\alpha$ is the fine-structure constant in the Thomson limit and $m_\pi$ and $m_\mu$ are the masses of the $\pi$ meson and $\mu$ lepton. The factor $K(t)/t$ in the integrand of (\ref{eq:amu}) exhibits a drastic increase at low $t$, which amplifies the weight of the poorly-known low-energy region in the integral. 

The advantage of the formulation (\ref{eq:amu}) is that one can exploit the analytic properties of the function $F_\pi^V(t)$ in the complex $t$ plane in order to perform its analytic continuation from other kinematical regions, where it is known with better precision, to the low-energy region of interest.
This approach has been followed in a series of recent papers \cite{Ananthanarayan:2012tn}-\cite{Ananthanarayan:2018nyx}, where we have explored the implications of analyticity and unitarity  on the pion form factor using methods based on functional analysis proposed in \cite{Caprini:1999ws, Abbas:2010jc}    (for a recent review see \cite{Caprini-Springer}).  In the present work we consider in more detail some aspects of this analysis and of its implications.

  After a brief  review of the method 
in Sec. \ref{sec:method}, we  compare  in Sec. \ref{sec:implic}  the form factor calculated at low $t$ by analytic continuation with the experimental determinations and other theoretical predictions available in this region. The sensitivity of the method to the input data  is explored in Sec. \ref{sec:input}, where we consider data from higher timelike energies, beyond the region used in the previous works. In Sec. \ref{sec:amu} we consider the implications on the low-energy contribution to $a_\mu$ and in Sec. \ref{sec:conc} we formulate our conclusions.
\section{Parametrization-free analytic continuation}\label{sec:method}
 Analyticity and unitarity provide powerful tools in hadron physics for performing the analytic continuation of scattering amplitudes and form factors  to energies where they are not precisely known.  In particular, the pion vector form factor is an analytic function  which satisfies the Schwarz reflection property $F_\pi^V(t^*)=(F_\pi^V(t))^*$ 
in the complex $t$ plane with a cut along  the real axis for $t\ge 4 m_\pi^2$. Along the cut the form factor is a complex function, written in terms of its modulus and phase  as
\beq\label{eq:Fpi}
F_\pi^V(t+i\epsilon)=|F_\pi^V(t)| e^{i\phi(t)},\quad\quad t\ge 4 m_\pi^2.
\eeq
According to Fermi-Watson theorem  \cite{Fermi:2008zz, Watson:1954uc}, below the first inelastic threshold the phase of the form factor is equal to the $P$-wave phase shift of $\pi\pi$ elastic scattering:
\beq\label{eq:watson}
\phi(t)=\delta_1^1(t),  \quad\quad  4 m_\pi^2 \le t \le t_{\text{in}},
\eeq
where one can take with a good approximation $\sqrt{\tin}=m_\omega+m_\pi=0.917\,\gev$.  In this region, the phase shift $\delta_1^1(t)$ is known with high precision  from Chiral Perturbation Theory (ChPT) and dispersion relations for $\pi\pi$ scattering \cite{Ananthanarayan:2000ht, Caprini:2011ky, GarciaMartin:2011cn}.

The standard dispersion theory for the pion form factor \cite{Barton, Oller} is  based on the Omn\`es representation,  which amounts to the reconstruction of an analytic function from its phase on the cut. However, this treatment requires  the knowledge of the phase above the inelastic threshold $\tin$ and the positions of the zeros in the complex plane, which are not known. Model-dependent assumptions on these quantities can be avoided by using above $\tin$,
 instead of the phase, the phenomenological  information available on the modulus from experimental measurements and perturbative QCD. Specifically, we adopt a conservative condition written as
\beq\label{eq:L2}
 \frac{1}{\pi} \int_{\tin}^{\infty}w(t)\, |F_\pi^V(t)|^2  dt \leq  I,
 \eeq
where $w(t)>0$ is a suitable weight for which the integral  converges and an accurate evaluation of $I$ from the available information is possible. 

One can use, in addition, several experimental values of the form factor on the spacelike axis:  
\beq\label{eq:val}
F_\pi^V(t_s)= F_s \pm \epsilon_s, \qquad t_s<0,
\eeq
and the   modulus measured at a finite number of points on the elastic region of the timelike axis
\beq\label{eq:mod}
|F_\pi^V(t_t)|= F_t \pm \epsilon_t, \qquad   4 m_\pi^2 < t_t <t_{\text{in}}.
\eeq

The conditions (\ref{eq:watson}) - (\ref{eq:mod}) cannot determine  the function $F_\pi^V(t)$ uniquely. However, using special techniques of functional analysis and optimization theory (for a review see \cite{Caprini-Springer}), one can derive rigorous upper and lower bounds on $F_\pi^V(t)$ for $t<4 m_\pi^2$ or the modulus $|F_\pi^V(t)|$ for $4 m_\pi^2<t<\tin$, in particular in the low-energy region of interest. The solution of the extremal problem, which is expressed by a positivity of a certain determinant, can be found in our previous works quoted above (see in particular Appendix A of Ref. \cite{Ananthanarayan:2018nyx}) and we shall not give it here.  For completeness, we shall briefly describe only the information used as input and the proper treatment of the statistical uncertainties.

\begin{figure}[thb]\begin{center}\vspace{0.5cm}
\includegraphics[width = 6cm]{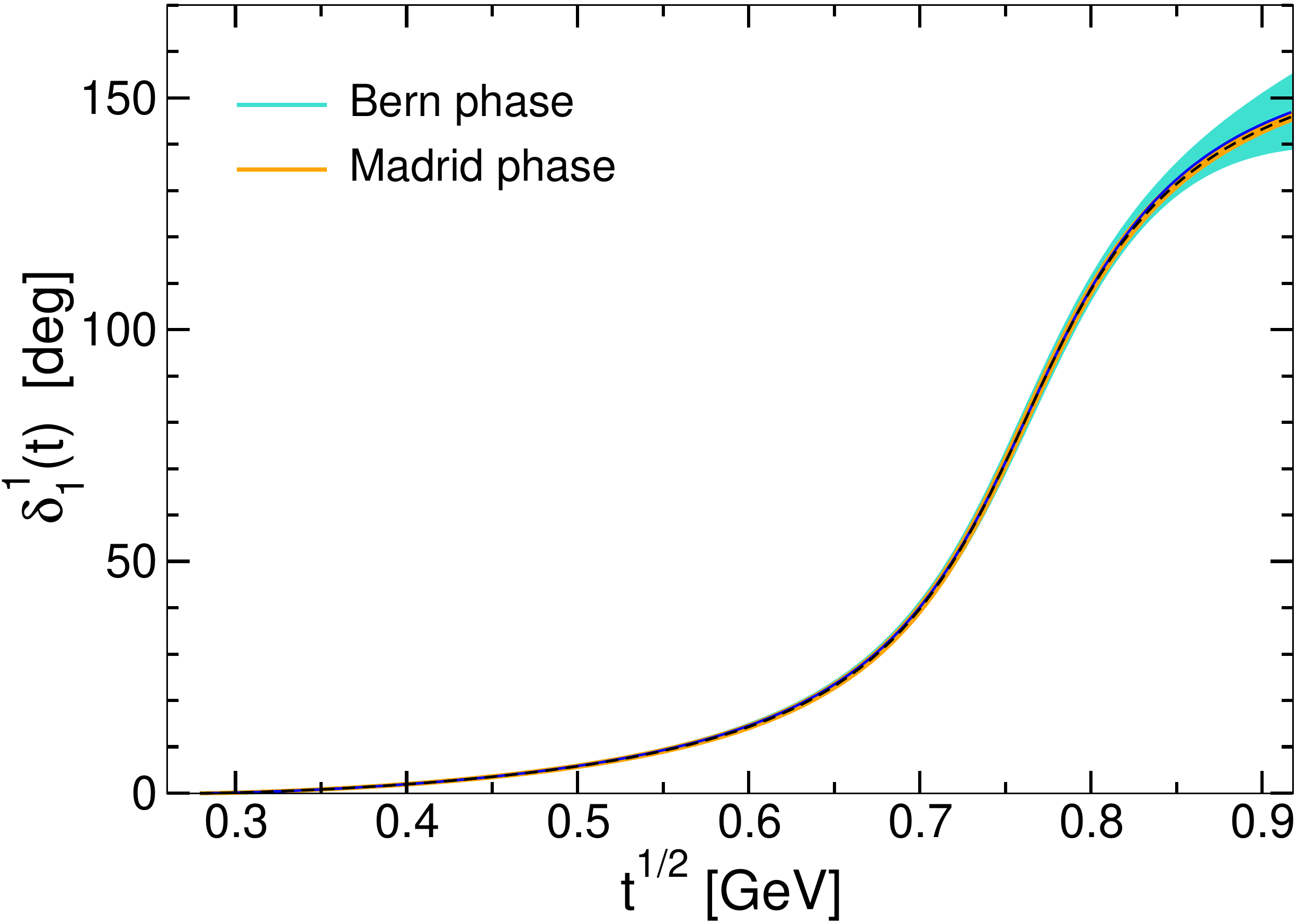}
\caption{The phase shift $\delta_1^1$ as a function of energy below the inelastic threshold $\tin=(m_\omega+m_\pi)^2$. The values calculated in \cite{Ananthanarayan:2000ht,Caprini:2011ky}   and \cite{GarciaMartin:2011cn} are denoted as Bern and Madrid phase, respectively.\label{fig:phases}}
\end{center}
\end{figure}
 As already mentioned above, the phase shift $\delta_1^1(t)$  appearing in  (\ref{eq:watson}) has been calculated with good precision  in Refs. \cite{Ananthanarayan:2000ht, Caprini:2011ky} and \cite{GarciaMartin:2011cn}, using ChPT and dispersion relations  for $\pi\pi$ scattering. The central values and the error bands of these phases,  which we denote as Bern and Madrid, respectively, are shown  in Fig. \ref{fig:phases}. One can note the remarkable consistency of the two solutions, with slightly larger uncertainties of the Bern phase  near $\tin$, which have actually little influence on the final results.

We have calculated the integral (\ref{eq:L2}) using the BABAR data \cite{Aubert:2009ad} 
from $\tin$ up to $\sqrt{t}=3\, \gev$, smoothly continued with  a constant value for the modulus 
in the range $3\, \gev \leq \sqrt{t} \leq 20\, \gev$,  and  a decrease  $\sim 1/t$ at higher energies, as predicted by perturbative QCD \cite{Farrar:1979aw,
Lepage:1979zb}. Finally, the input in (\ref{eq:val}) and  (\ref{eq:mod}) was taken from
the most recent experimental measurements of  $F_\pi$ Collaboration at JLab \cite{Horn, Huber}, and the modulus measured by the $e^+e^-$ experiments \cite{Akhmetshin:2003zn}-\cite{Ablikim:2015orh} in the region (0.65 - 0.71) GeV,  which we denoted as ``stability region'' because the  data have here good precision  and the determinations of different   experiments are consistent.

 A nontrivial complication is the fact that the experimental values used as input are beset by statistical errors. 
This requires to properly merge the formalism of analytic bounds with statistical simulations.
The problem was solved in Refs. \cite{Ananthanarayan:2016mns, Ananthanarayan:2017efc, Ananthanarayan:2018nyx} by generating a large sample of pseudo-data, 
achieved by randomly sampling each of the
input quantities with specific distributions based on the measured central values and the quoted errors.
For each point from the input statistical sample,   upper and lower bounds on $F_\pi^V(t)$ (or $|F_\pi^V(t)|$) at points $t$ in the kinematical regions of interest have been calculated using the formalism described above.   Finally, a set of values  in between the bounds has been uniformly generated, taking into account the fact that all the values between the extreme points are equally valid. 

In this way, for each spacelike and timelike input  a large  sample of output values of the form factor (or its modulus) at points $t$ of interest have been generated.  The output distributions turn out to be close to a Gaussian, allowing the extraction of the mean value and the standard deviation (defined as the 68.3\% confidence limit interval). The values obtained with input from different energies  and different experiments have been then combined using a prescription proposed in \cite{Schmelling:1994pz}, where the correlation between different values is derived from the values themselves. The procedure has been performed separately with Bern and Madrid phases, the  average of the corresponding values being adopted as final result. 

\section{Analytic continuation to low $t$ versus direct determinations}\label{sec:implic}
\begin{figure*}[thb]\hspace{1.cm}\includegraphics[width = 6.45cm]{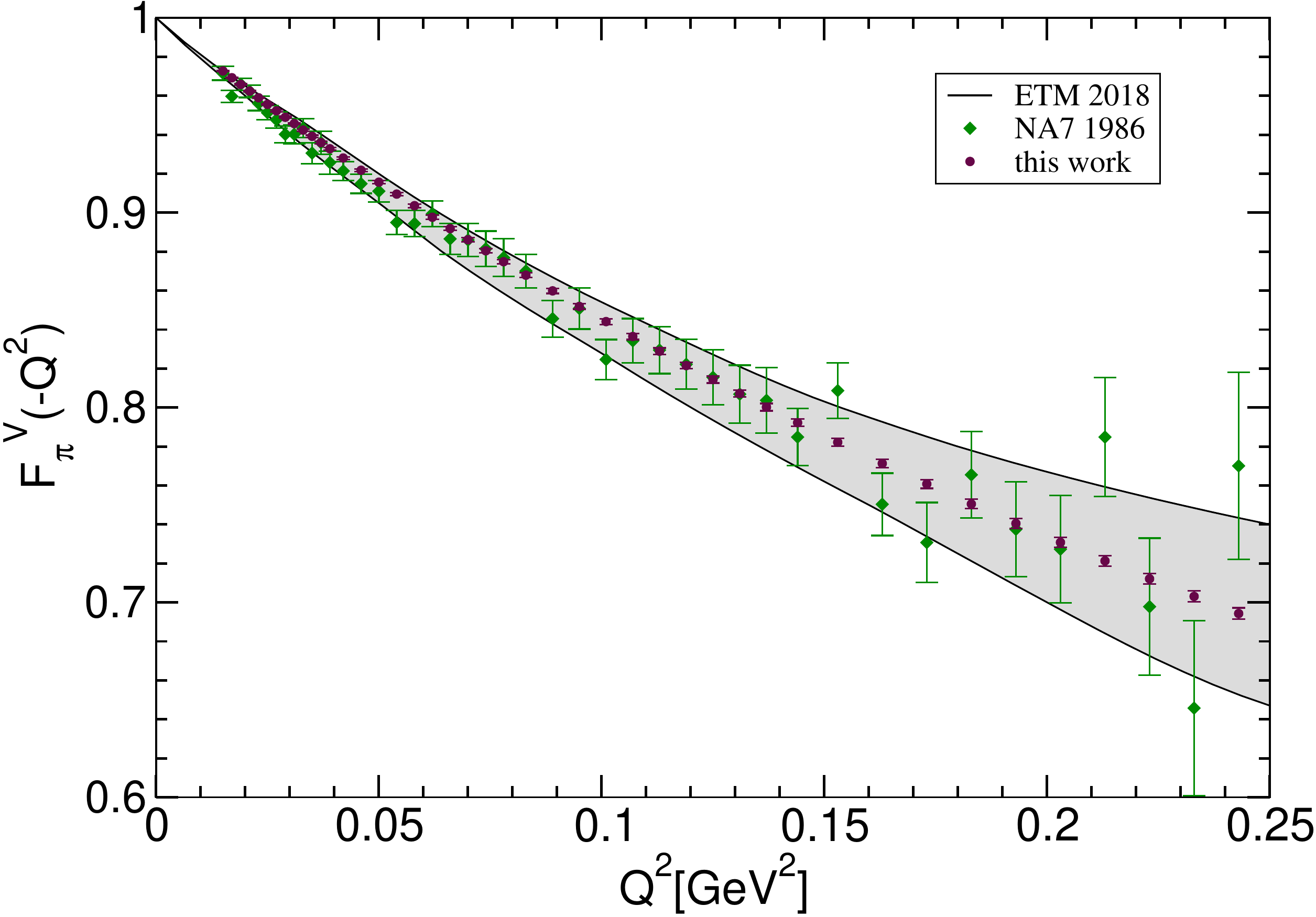}\hspace{1.2cm}
\includegraphics[width = 6.cm]{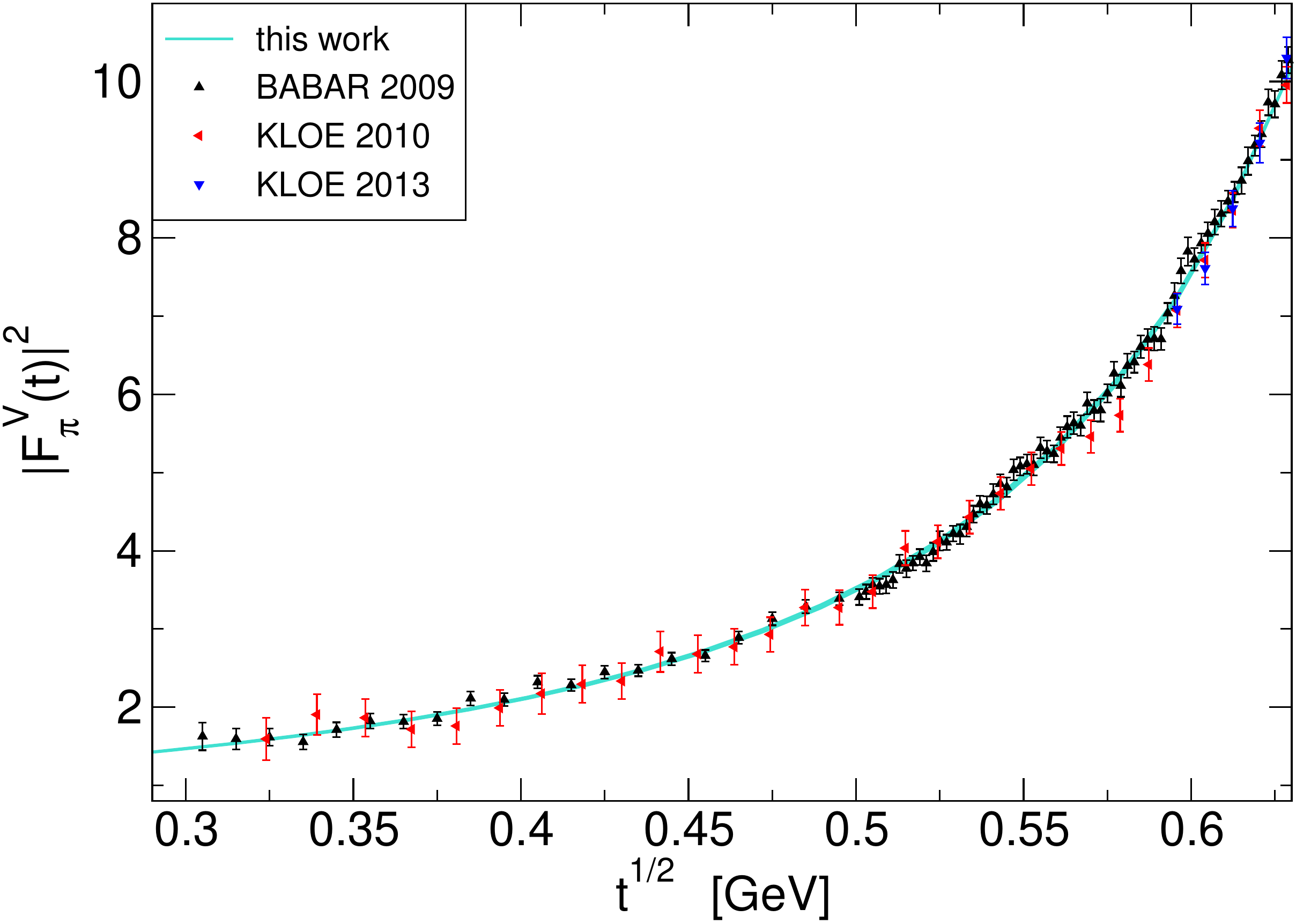}	
		\caption{Left: pion electromagnetic form factor in the spacelike region near the origin, compared with experimental data \cite{Amendolia:1986wj} and lattice QCD calculations \cite{ETM}.  Right:  modulus of the form factor on the timelike axis below $0.63\, \gev$, compared with BABAR \cite{Aubert:2009ad} and KLOE \cite{Ambrosino:2010bv, Babusci:2012rp} data.\label{fig:NA7}}
\end{figure*}

 It is of interest to confront the values obtained by  analytic continuation with the direct determinations from experiment or theory at low momenta. The comparison performed in \cite{Ananthanarayan:2018nyx} is illustrated in Fig. \ref{fig:NA7}, where the extrapolated values are shown together with the measurements  of NA7 Collaboration \cite{Amendolia:1986wj} and the lattice calculations of ETM Collaboration \cite{ETM} at low spacelike values $Q^2=-t>0$ (left panel), and with the measurements of BABAR \cite{Aubert:2009ad} and KLOE \cite{Ambrosino:2010bv, Babusci:2012rp} in the timelike range $4 m_\pi^2<t\leq (0.63 \gev)^2$ (right panel).
 Fig. \ref{fig:NA7} shows that the values obtained by analytic extrapolation have smaller errors and  are in agreement with the available  data.

 For assessing in a quantitative way the agreement we shall use the quantity 
\begin{widetext}\beq\label{eq:chi2}
\chi^2=\sum_{i,j=1}^{n}(F_{\text{extrap}}(x_i)-F_{\text{direct}}(x_i))(\text{Cov}^{-1})_{ij} (F_{\text{extrap}}(x_j)-F_{\text{direct}}(x_j)),
\eeq\end{widetext}
where the sum is over the points where direct data are available and $\text{Cov}$ is the full covariance matrix, assumed first to be diagonal by neglecting the correlations between data points.

For the spacelike points we obtained $\chi^2/n= 1.34$ for experimental data  and   $\chi^2/n= 0.09$ for lattice QCD data. The small value of  $\chi^2/n$ for lattice QCD is explained by the fact that these calculations have still large uncertainties, as seen from Fig. \ref{fig:NA7}.

For $t$ on the timelike axis,  we
obtained for BABAR data the value $\chi^2/n=0.82$ using only the experimental errors and $\chi^2/n= 0.79$ by combining in quadrature at each point
the experimental error and the theoretical error of the extrapolated values. For KLOE experiment \cite{Ambrosino:2010bv},  we obtained $\chi^2/n=0.56$ using only the experimental errors and $\chi^2/n=0.54$ with total errors.  The smaller values of $\chi^2/n$  are due to the larger errors of KLOE data, seen also in Fig. \ref{fig:NA7}. 
 The inclusion of the correlations is expected to change only slightly these values. For instance, using the BABAR covariance matrix from \cite{Aubert:2009ad}, the results quoted above for BABAR become 0.81 and 0.77, respectively.
 We conclude that the data available at low $t$ are consistent with the form factor calculated by analytic extrapolation in a model-independent formalism that exploits analyticity and unitarity.

In the analysis reported above the timelike input (\ref{eq:mod}) was restricted to the region (0.65 - 0.71) GeV.
 In the next two sections we will explore the sensitivity of the method to the input region and the implications on the muon $g-2$.

\section{Choice of the timelike input}\label{sec:input}
As mentioned above, in the previous works \cite{Ananthanarayan:2016mns, Ananthanarayan:2017efc, Ananthanarayan:2018nyx} the timelike input (\ref{eq:mod}) was taken from the ``stability region'' (0.65 - 0.71) GeV,  where the  determinations   of different   experiments have good precision and  are consistent. Another argument in favour of  this region is its proximity to the range $\sqrt{t}<0.63 \gev$, where the extrapolation is performed.

\begin{figure*}[thb]\label{fig:hist}
	\begin{center}
		\includegraphics[width = 6cm]{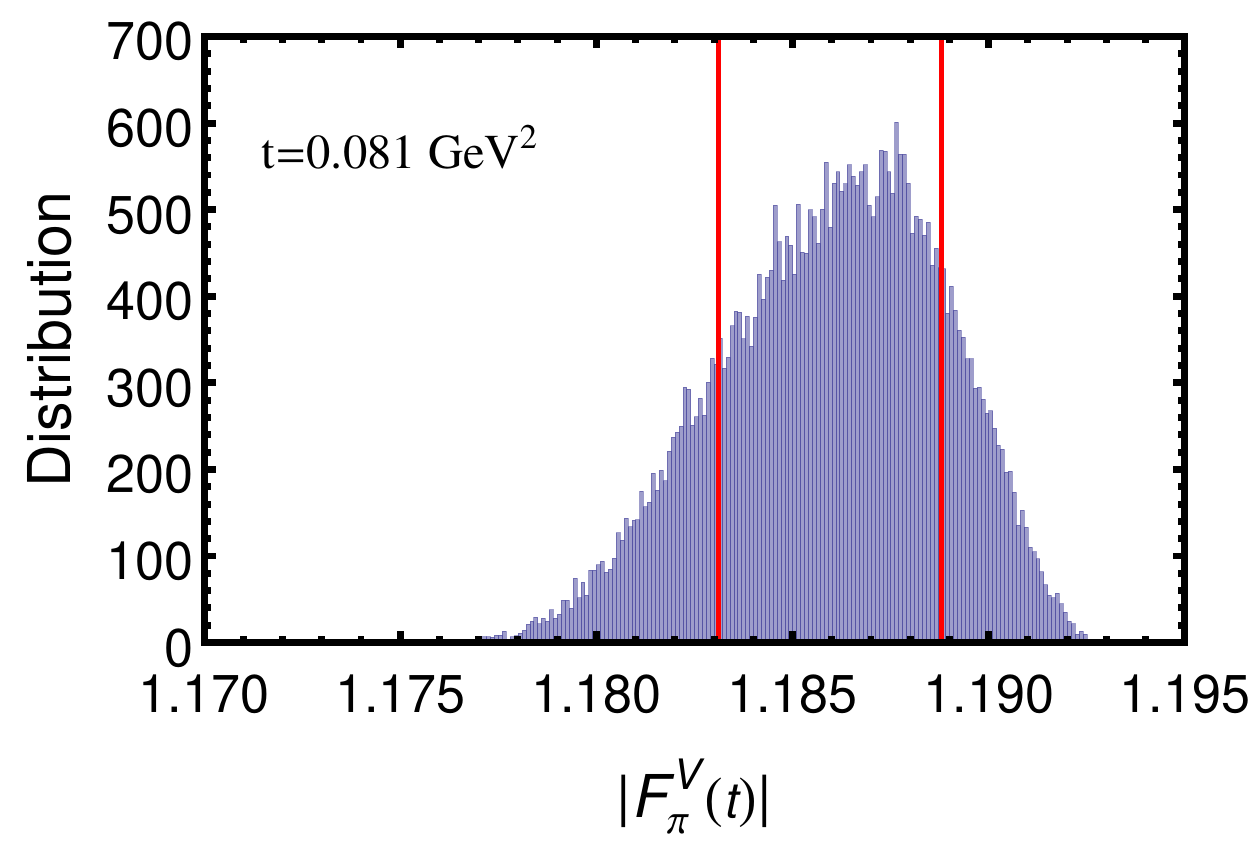}\hspace{1.5cm}
		\includegraphics[width = 6cm]{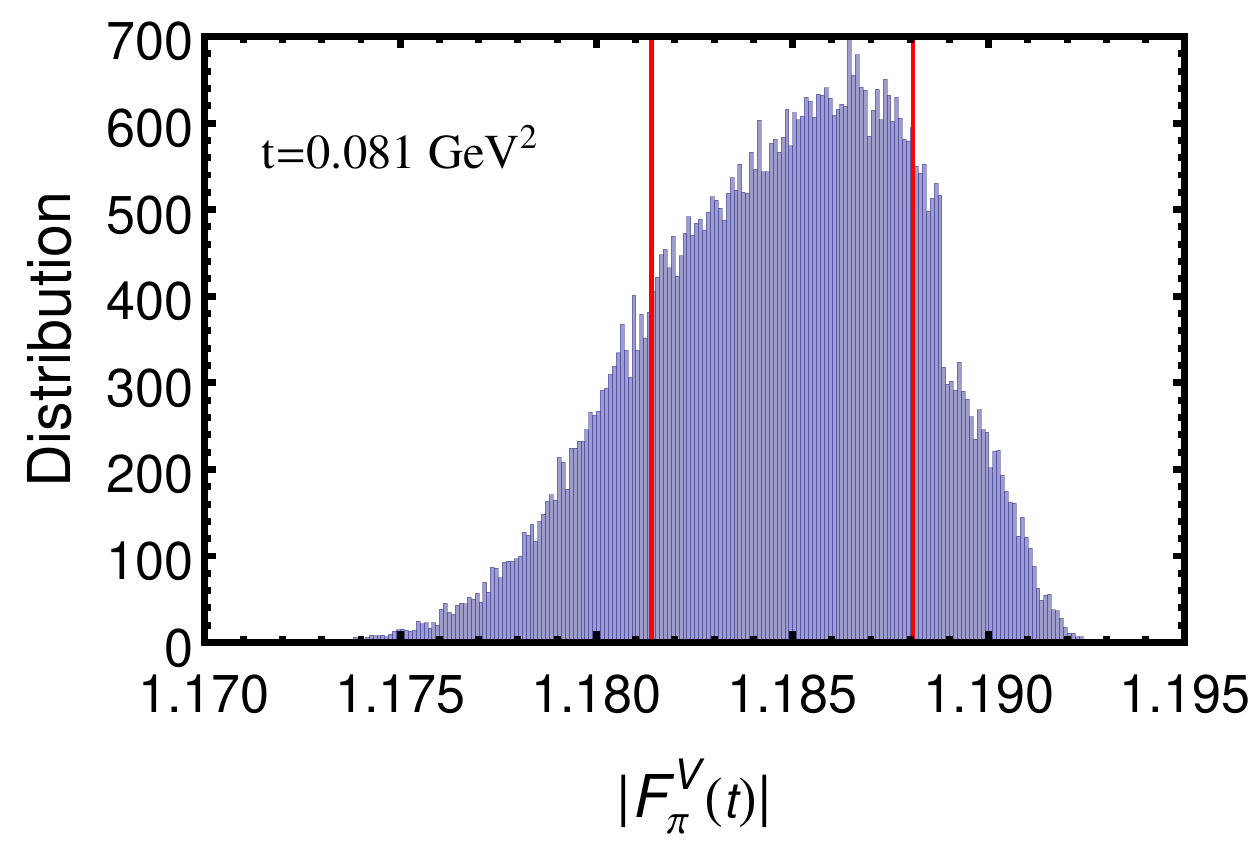}\\\includegraphics[width = 5cm]{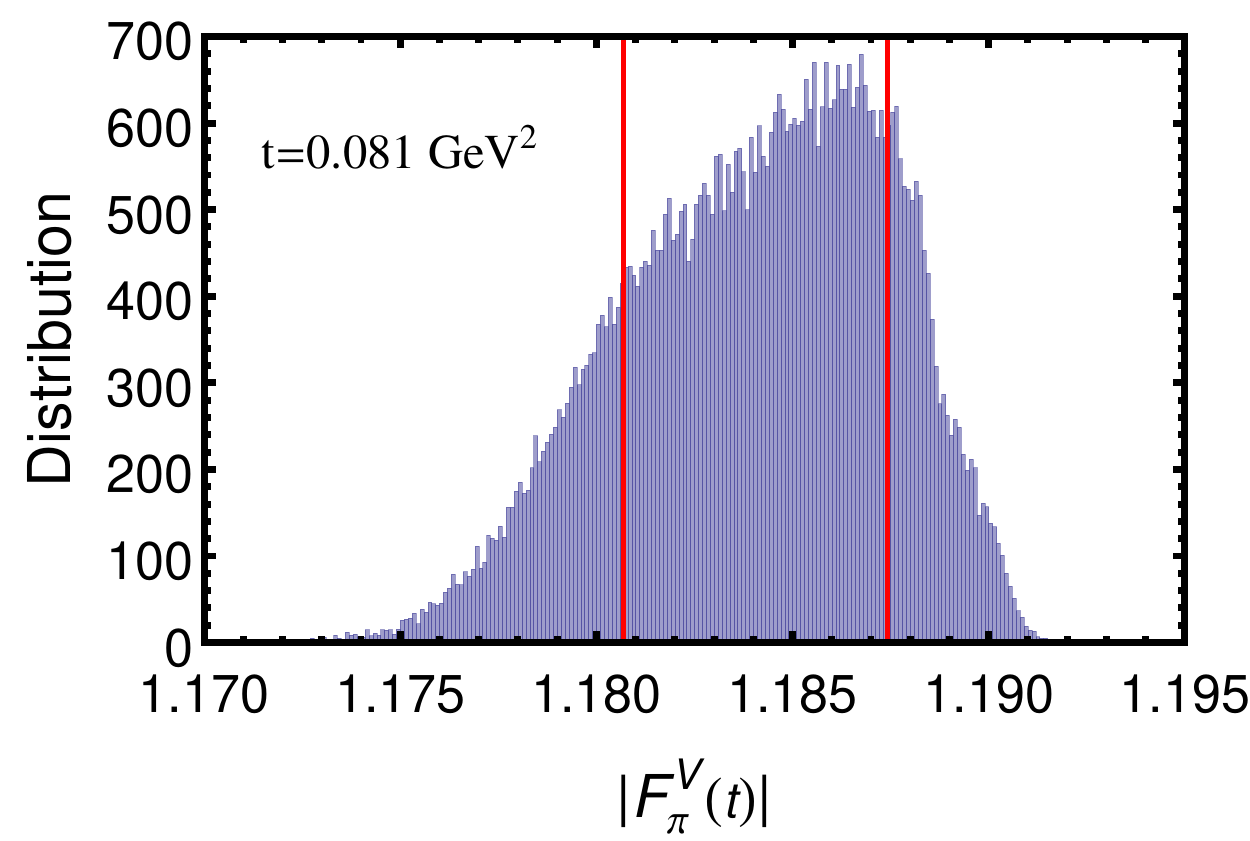}\hspace{1.5cm}
		\includegraphics[width = 6cm]{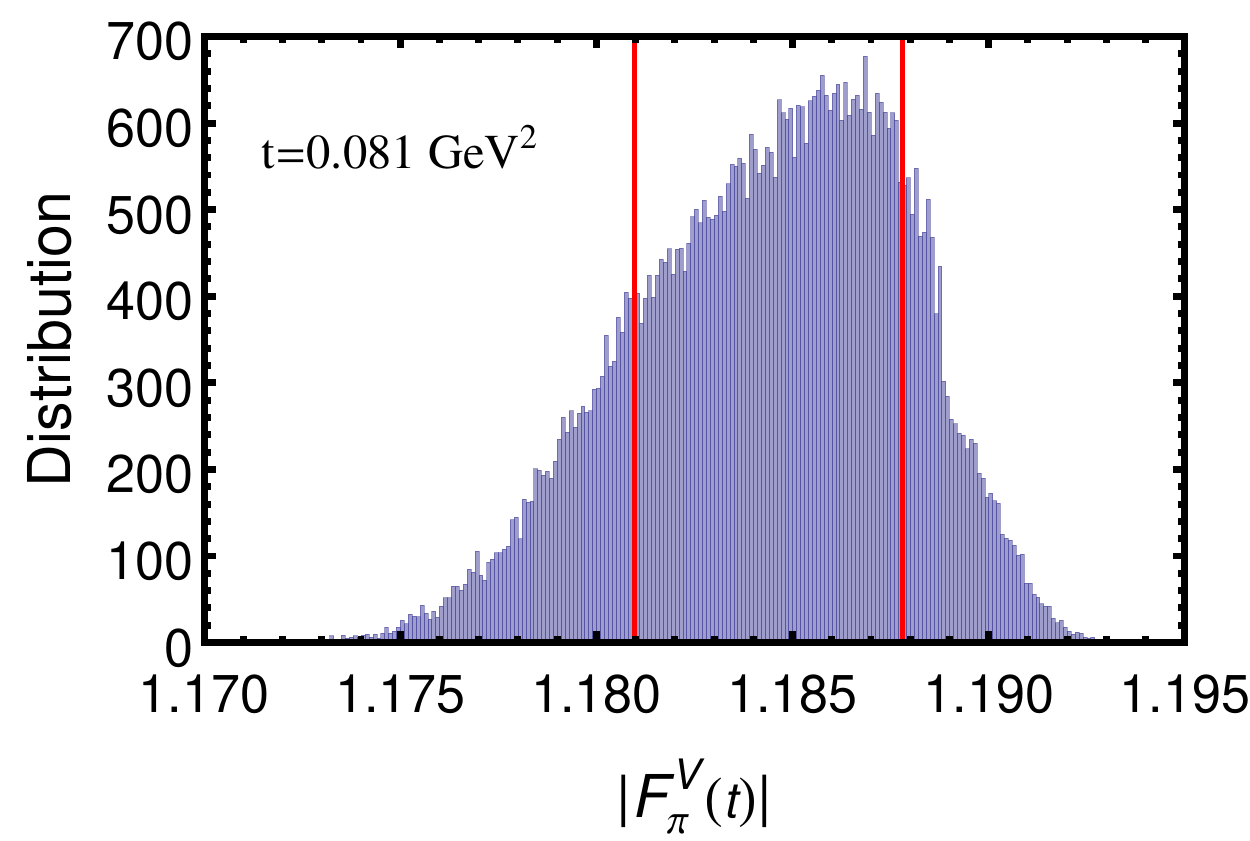}
		\caption{Statistical distributions of the values of $|F_\pi^V(t)|$  at $t=0.081 \gev^2$, with input modulus at  $\sqrt{t_t}= 0.699  \gev$ 
from \cite{Aubert:2009ad} (first panel) and   $\sqrt{t_t}= 0.76  \gev$ from Refs. \cite{Aubert:2009ad, Ambrosino:2010bv, Babusci:2012rp} (the remaining three panels). The red vertical lines
indicate the 68.3 \% confidence limit (CL) intervals.
 }
	\end{center}
\end{figure*}

  It may be noted however that this choice is an educated guess and cannot necessarily be
considered very rigorous. Discrepancies between different experiments  are indeed present at higher energies: they can be inferred from Fig. 4 of Ref. \cite{Teubner:2018}, which shows that the contributions to $a_\mu$ from the range (0.6 - 0.9) GeV calculated with BABAR and KLOE data are significantly different. However, from the data compilation presented in Fig. 6 of the quoted paper one may see that the discrepancies start to matter only at energies above  0.76 GeV, becoming significant especially above the peak of the $\rho$ resonance.

 Therefore, it is of interest to explore, in the framework of the method presented in Sec. \ref{sec:method}, the effect of a timelike input (\ref{eq:mod})  from a higher energy. For illustration, we show in Fig. \ref{fig:hist} the statistical distributions of the output modulus of the  form factor at $t=0.081\,  \gev^2$,  using input modulus at  $\sqrt{t_t}= 0.699\,  \gev$ 
from BABAR (Ref. \cite{Aubert:2009ad}) in the first panel, and   $\sqrt{t_t}= 0.76\,  \gev$ from BABAR, KLOE 2010 and KLOE 2013 (Refs. \cite{Aubert:2009ad, Ambrosino:2010bv, Babusci:2012rp}) in the remaining three panels.

One can see that the standard deviations, obtained from the 68.3\% CL intervals, are slightly larger for the timelike input from the higher energy, as may be expected for the extrapolation from more distant points. However, the increase of the uncertainty is not dramatic and the mean values are mutually consistent. We conclude that the input timelike region  can be extended up to about 0.76 GeV. While the benefit for extrapolations to low values of $t$ may be not significant (as noticed in practice,  additional points do not reduce automatically the uncertainty of the combined value), the extended input is expected to improve the extrapolation to higher  $t$, in particular above the $\rho$ peak. This problem will be investigated in a future work.

\section{Implications for muon $g-2$}\label{sec:amu}
Several  determinations of the contribution to $a_\mu$ of the hadronic vacuum polarization from energies below 0.63 GeV have been reported recently.
The direct integration of a compilation of the  $e^+e^-$ cross-section data, performed  in \cite{Davier:2017}, gives
\beq\label{eq:Dav}
a_\mu^{\pi\pi}|_{\leq 0.63 \gev} = (133.12 \pm 1.31)\times 10^{-10},
\eeq
while the interpolation performed in \cite{Teubner:2018} leads to\footnote{We thank T. Teubner for sending us this value.}
\beq\label{eq:Teub}
a_\mu^{\pi\pi}|_{\leq 0.63 \gev} = (131.12 \pm 1.03)\times 10^{-10}.
\eeq
We quote also the result 
\beq\label{eq:colang}
 a_\mu^{\pi\pi}|_{\leq 0.63 \gev} =(132.5\pm 1.1)\times 10^{-10}
\eeq  of the recent analysis \cite{Colangelo:2018mtw}, which exploits  analyticity and unitarity by using an extended Omn\`es representation of the pion form factor in a global fit of the data on $e^+e^-\to \pi^+\pi^-$ cross section below 1 GeV and the NA7 experiment \cite{Amendolia:1986wj}.

For comparison, the method described in Sec. \ref{sec:method} with input (\ref{eq:mod}) on the modulus of the pion form factor in the range (0.65 - 0.71) GeV from all the $e^+e^-$ experiments leads to the prediction \cite{Ananthanarayan:2018nyx}:
\beq\label{eq:res1}
a_\mu^{\pi\pi}|_{\leq 0.63 \gev} = (132.91\pm 0.76)\times 10^{-10}.
\eeq
This result is consistent with the other determinations and has a better precision.

In order to assess the sensitivity to the input, we have performed the calculation of $a_\mu^{\pi\pi}|_{\leq 0.63 \gev}$ with the same method, but using as  input the modulus of the form factor measured  in the region (0.70 - 0.76) GeV. For a detailed comparison, we give the values obtained with data from separate experiments: $(133.62 \pm 0.67)\times 10^{-10}$ using BABAR data \cite{Aubert:2009ad}, $(132.43 \pm 1.31)\times 10^{-10}$ using KLOE 2010 data \cite{Ambrosino:2010bv} and $(132.31 \pm 1.45)\times 10^{-10}$ using KLOE 2013 data \cite{Babusci:2012rp}.
These values are consistent within errors. Moreover, they are consistent with our prediction (\ref{eq:res1}), obtained with  input on the modulus from the range (0.65 - 0.71) GeV. This proves the robustness of the method and the stability of the results towards the change of the input.  

\section{Summary and discussion}\label{sec:conc}
 The  theoretical  evaluation  of the muon magnetic anomaly $a_\mu$  in the Standard Model requires the precise knowledge of the modulus of the pion electromagnetic form factor $F_\pi^V(t)$ at low values of $t$ on the timelike axis. However, the experimental dtermination of this quantity at low energies is difficult.

 As shown in a series of recent papers \cite{Ananthanarayan:2016mns, Ananthanarayan:2017efc, Ananthanarayan:2018nyx}, it is possible to obtain the form factor at low momenta  by analytic continuation from regions where it is more precisely measured, implementing also the phase known from $\pi\pi$ scattering via Fermi-Watson theorem.  The method proposed in these works does not rely on specific parametrizations, using instead the solution of an extremal problem combined with statistical simulations. 

 In the present paper we have performed several consistency and sensitivity tests of the method. First, in Sec. \ref{sec:implic} we compared in a quantitative way the results obtained by analytic extrapolation with the data available at low momenta from experiment or from lattice QCD.  The good agreement of these values provides a precise test of the consistency of the data with analyticity and unitarity. 

We have then explored the sensitivity of the method to the information on the modulus used as input, by considering the region (0.70 - 0.76) GeV, higher than the stability region (0.65 - 0.71) GeV used in the previous works.  The results presented in Sec. \ref{sec:input}   prove the robustness of the method and its stability to the variation of the input. The stability is confirmed also by the values of $a_\mu^{\pi\pi}|_{\leq 0.63 \gev}$ given in Sec. \ref{sec:amu}.

 The detailed influence of the extended input region on the extrapolation to low energies and the challenging extrapolation to energies above the $\rho$ peak, where significant discrepancies among the experiments exist, will be investigated in a future work.

\subsection*{Acknowledgments} We thank Michel Davier  for an interesting correspondence which stimulated us to perform this work. I.C. acknowledges support from the   Ministry of Research and Innovation of Romania, Contract PN 18090101/2018. D.D. is supported by the DST, Government of India, under INSPIRE Fellowship
(No. DST/INSPIRE/04/2016/002620).


\end{document}